\documentclass[article,prd]{revtex4}

\usepackage{amsbsy}
\usepackage{amssymb}
\usepackage{amsmath}
\usepackage{graphicx}

\usepackage[british]{babel}

\usepackage{color}

\usepackage{cancel}

\def\X{{\mathrm{x}}}

\def\Y{{\mathrm{y}}}
\def\x{{\mathrm{x}}}
\def\y{{\mathrm{y}}}

\def\n{{\mathrm{n}}}

\def\s{{\mathrm{s}}}

\def\k{{\rm k}}
\def\n{{\rm n}}
\def\p{{\rm p}}
\def\e{{\rm e}}

\def\s{\mathrm{s}}

\newcommand{\A}{{\cal A}}
\newcommand{\B}{{\cal B}}

\def\n{{\rm n}}
\def\p{{\rm p}}
\def\e{{\rm e}}

\def\B{{\mathcal B}}
\def\A{{\mathcal A}}

\def\e{{\rm e}}
\def\s{{\rm s}}

\def\be{\begin{equation}}
\def\ee{\end{equation}}
\def\beq{\begin{equation}}
\def\eeq{\end{equation}}
\def\bea{\begin{eqnarray}}
\def\eea{\end{eqnarray}}

\def\bear{\begin{eqnarray}}
\def\eear{\end{eqnarray}}

\begin{document}

\title{A multifluid perspective on multimessenger modelling}

\author{N. Andersson}

\affiliation{Mathematical Sciences and STAG Research Centre, University of Southampton,
Southampton SO17 1BJ, United Kingdom}

\begin{abstract}
 This brief review introduces the notion of a relativistic multifluid system---a multi-component system with identifiable relative flows---and outlines a set of models for scenarios relevant for different astronomical observation channels. The specific problems used to illustrate the key principles include superfluid hydrodynamics (with relevance for radio and x-ray pulsar timing and gravitational-wave searches), heat flow (connecting to the problem of neutron star cooling and associated x-ray observations) and the coupling between matter and electromagnetism (linking to explosive phenomena like gamma-ray bursts and more subtle issues like the long-term evolution of a neutron star's magnetic field). We also comment on the coupling between matter and radiation, for which the multifluid approach would seem less appropriate. The main motivation of the survey is to illustrate less familiar aspects that come into play in multifluid problems, establish the relevant ``language'' and provide a platform for more detailed work on these issues.
\end{abstract}

\maketitle

\section{Multi-messenger modelling}

The spectacular neutron star merger observation on 17 August 2017 \cite{2017ApJ...848L..12A} provided a number of firsts for astronomy: The direct detection of gravitational waves from systems involving neutron stars. Indisputable support for the connection between neutron star mergers and short gamma-ray bursts \cite{2017ApJ...848L..13A}. Clear evidence for  mass outflows leading to rapid nuclear reactions and kilonova emission \cite{2017ApJ...848L..17C,2017ApJ...848L..21A}. A meaningful constraint  on the neutron star radius via the tidal deformability imprint (or perhaps rather, lack of) on the inspiral signal \cite{2017ApJ...848L..12A,2018PhRvL.121i1102D}. With astronomers tracking the event across the electromagnetic spectrum, while gravitational-wave analysts tried to tease out as much information from the signal as possible,  this was an awesome opening episode\footnote{From a historical perspective, one might want to argue that SN1987A represents a (much) earlier example of multi-messenger astronomy, with neutrinos associated with the event observed along with the visible light \cite{1987PhRvL..58.1490H}. However, it is equally fair to argue that, in that case the different observation channels did not provide comparable ``high-quality'' information.} for the era of multi-messenger astronomy. 

As we sober up from the celebrations, it is natural to consider what we have to do---as theorists---to keep up with the rapidly improving precision of the observations---driven by to the seemingly inexhaustible ingenuity of the teams that design new and improved instruments. One thing is clear: We may soon reach the point where  data extracted from observations require theoretical models with a precision  we are (at this point) unable to provide. This is particularly the case for problems involving neutron stars.
These are complex systems, the description of which requires much of the physics we know (and some aspects we do not know particularly well, if at all). The issues range from nuclear physics (both the strong and weak interaction), to gravity, electromagnetism and features from condensed matter and low-temperature physics \cite{2017IJMPD..2630015G}, as well. In essence, the modelling of these systems is a playground for theoretical physics. This is inspiring and intimidating in---possibly depending on personality  and inclination---equal measure.

Much of what we (think we) know about neutron stars build on relatively simple models of hydrostatic equilibrium, essentially solutions to the Tolman-Oppenheimer-Volkoff equations for some given matter equation of state. This allows us to consider issues like the maximum mass \cite{2020ApJ...901..155G} (above which the star must collapse to form a black hole) and the range of expected radii \cite{2019ApJ...887L..22R,2019ApJ...887L..24M}, both of which have been confronted by observations. Dynamical problems are, inevitably, more complicated. We have to consider how high-density matter moves and neutron stars involve a number of (at least to some extent) distinguishable matter components. In fact, the simplest ``reasonable'' neutron star model involves four distinct fluid components \cite{2017CQGra..34l5002A} (although one may argue that we should consider as many as seven components \cite{2020PhRvD.102f3011R}, which could be (way) more than we can realistically handle). Heat may flow relative to matter. Charges have to flow relative to one another to support the star's large-scale magnetic field. Superfluid components may be fairly weakly coupled to the rest of the star. These different features are independent of one another (although coupled) so if we also consider the bulk matter flow we end up with four distinct ``fluxes''.
This realisation takes us beyond the familiar territory of text-book fluid dynamics into less frequently travelled terrain. An aspect that comes to the fore is the ability of different fluid components (sharing the same volume) to flow---or perhaps closer to reality, drift--relative to each other. We are dealing with a multifluid system and we need to understand what this implies.
The issues involved are both conceptual and practical.

The desire to explore this kind of system motivates this brief overview. Drawing on a recent exhaustive review\footnote{As the material in \cite{2007LRR....10....1A} provides much of the relevant context material and links to the relevant literature, we have not tried to be all-inclusive here. The aim is to provide an introduction and a starting point rather than an encyclopedic review.} of relativistic fluid dynamics \cite{2007LRR....10....1A}, we consider some of the issues that arise when we approach the dynamics of multifluid systems, highlight new aspects that arise and sketch how we may be able to build the next generation of models for relevant neutron star scenarios. 

\section{multifluid models} 

The convective variational approach pioneered by Carter  \cite{carter89:_covar_theor_conduc} and a number of collaborators (see \cite{2007LRR....10....1A} for a detailed review) provides a natural framework for a discussion of relativistic multifluid systems. The approach is fairly intuitive. Consider a system with a number of identifiable matter constituents, labelled by $\x,\y,\ldots$, each associated with a flux 
\be
n_\x^a = n_\x u_\x^a \ ,
\ee
(we use $a,b,c, \ldots$ to represent spacetime indices and it should be noted that the label $\x$ is not summed over when repeated). Each four velocity is normalised, $g_{ab} u_\x^a u_\x^b = - 1$ (with $g_{ab}$ the spacetime metric and we assume units such that $c=G=1$ throughout the discussion), and the number density measured by a co-moving observer is
\be
n_\x = - u^\x_a n_\x^a \ .
\ee
 
The variational principle involves a matter Lagrangian, $\Lambda$, taken to be a relativistic 
invariant and hence depending only on covariant scalars formed out of the different fluxes. Effectively, this provides the equation of state for matter. In the general case, we have to consider both 
\be
n_\x^2 = - g_{ab} n_\x^a n_\x^b \ ,
\ee
and
\be
n_{\x\y}^2 = - g_{ab}n_\x^a n_\y^b \quad  \mbox{with}\quad  \y \neq \x \ .
\ee
An unconstrained variation of $\Lambda$ with respect to the fluxes $n^a_\x$ and the spacetime metric 
then gives (ignoring terms that can be written as total derivatives, which may be thought of as ``surface terms'' in the action)
\beq
    \delta \left(\sqrt{- g} \Lambda\right) = \sqrt{- g} \left[\sum_{\x} 
    \mu^\x_a \delta n^a_\x + \frac{1}{2} \left(\Lambda g^{a b} + \sum_{\x} 
    n^a_\x \mu^b_\x\right) \delta g_{a b}\right] \ , \label{dlamb}
\eeq
where $g$ is the determinant of the metric. This defines the individual fluid momenta 
\beq
     \mu^\X_{a} = \left( {\partial \Lambda \over \partial n_\x^a} \right)_{n_\y^b} =  g_{ab} \left(\B^{\X } n^b_\X + \sum_{\y\neq \x} \A^{\X \Y} 
                   n^b_\y\right) \ , 
                   \label{momdef}
                   \end{equation} 
with
\begin{equation}
\B^\x =  - 2 \frac{\partial \Lambda}{\partial 
                 n^2_{\X}} \ ,
\eeq
 and
\beq
                  \A^{\X \Y} = \A^{\Y \X} = - \frac{\partial \Lambda}{\partial 
                 n^2_{\X \Y}} \quad , \quad \y \neq \x \ . \label{coef12} 
\end{equation} 
The  covector $\mu^\X_a$  provides the fluid momentum and its magnitude gives
the chemical potential (intuitively, the energy associated with adding a particle of the $\x$-species to the system). For a co-moving observer, we have
\be
\mu_\x = - u_\x^a \mu^\x_a \ .
\ee
We can now identify one of the key features of a multifluid model. Each momentum may be ``tilted'' relative to the corresponding matter current. This is the so-called 
entrainment effect \cite{2007LRR....10....1A,andreev75:_entrain,comer03:_rel_ent}, and  it enters the model through the $\A^{\x\y}$ coefficients. We will consider the relevance of this  later.

Returning to the variational argument, equation \eqref{dlamb} shows why it is necessary to constrain the action. As it stands, the variation of $\Lambda$ suggests that the equations of motion 
would be $\mu^\x_a=0$---none of the fluids carry  energy or momentum, which is not a very interesting situation to consider. The actual equations of motion follow if we constrain the fluxes to be conserved\footnote{The convective variational approach builds on the assumption of conserved fluxes. Strictly, this means that we need to amend the strategy if we want to consider non-conserved flows, as in the case of dissipative systems. We will sketch some results in this direction, focussing on ``resistivity'' later. The general problem is much more complicated \cite{2015CQGra..32g5008A,2020arXiv200800945C}. }:
\begin{equation}
    \nabla_a n_\x^a = 0 \ .
    \label{divn}
\end{equation}
This is typically done  by changing to a Lagrangian perspective and making use of a lower-dimensional matter space \cite{carter89:_covar_theor_conduc}. The relevant steps are outlined in \cite{2007LRR....10....1A} and we will not repeat them here. Our focus will be on applying the model rather than the derivation. The constrained variation leads to 
\beq
    \delta \left(\sqrt{- g} \Lambda\right) =  \sqrt{- g} \left[ 
   \frac{1}{2} \left(\Psi \delta^a{}_b + \sum_{\X } n^{a}_\X 
    \mu_b^\X\right) g^{b c} \delta g_{a c} -  
   \sum_{\X } f^\X_{a} \xi^{a}_\X\right]
    \ , \label{variable}
\eeq
where $\xi_\x^a$ represents the Lagrangian displacement associated with each fluid's set of worldlines, which are taken to be conserved as a consequence of \eqref{divn}. We have also 
introduced the ``force'' acting on the x-component, $f_\x^a$, leading to the equations of motion 
\beq 
     f^\X_b \equiv  2 n^a_\X  \nabla_{[a} \mu^\x_{b]} =0 \ ,
     \label{forcex}
\eeq
(since the displacements are independent and arbitrary).
It also follows that  the stress-energy tensor (the variation with respect to the spacetime metric) takes the form
\begin{equation} 
     T^{a}{}_{b} = \Psi \delta^a{}_b + \sum_{\X} 
                       n^a_\X \mu^\X_b \ , 
\label{tab}\eeq
where
\beq
    \Psi = \Lambda - \sum_{\X} n^{a}_\X \mu^\X_{a} \ ,   \label{presdef} 
\end{equation} 
is a  generalised pressure. We will make this notion more precise in the following.

The equations we have written down may seem somewhat peculiar. The standard approach to relativistic fluid dynamics takes the divergence of the stress-energy tensor---and the fact that it must vanish by virtue of the Einstein equations and the Bianchi identities---as the starting point. Here we seem to instead have a set of force-balance relations represented by \eqref{forcex}, involving the vorticity formed from each fluid momentum
\begin{equation}
\omega^\x_{ab} = 2 \nabla_{[a}\mu^\x_{b]}\ .
\label{vort}
\end{equation}
However, the two pictures are perfectly consistent. We have
\begin{multline}
    \nabla_a T^{a}_{\ b} = \nabla_b \Lambda - \nabla_b \left( \sum_{\X} n^{a}_\X \mu^\X_{a} \right) + \nabla_a \left(  \sum_{\X} 
                       n^a_\X \mu^\X_b \right) \\
= \underbrace{  \nabla_b \Lambda - \sum_{\X}  \mu^\X_{a} \nabla_b n^{a}_\X}_{=0} - \sum_\x n_\x^a \nabla_b \mu^\x_a + \sum_\x \mu_b^\x \underbrace{ \nabla_a n_\x^a}_{=0} + \sum_\x n^a_\x \nabla_a \mu^\x_b \\
= \sum_\x f^\x_b = 0 \ .
\label{divT}
\end{multline}
In other words, if
 the  set of  equations \eqref{forcex} are satisfied 
then $\nabla_{a} T^{a}{}_{b} = 0$ follows as an identify. Intuitively, the argument in \eqref{divT} is ``Newton's third law'' in action.

Let us keep the focus on the stress-energy tensor \eqref{tab}. It may be helpful to, first of all, highlight the single-fluid result (simply dropping the chemical labels as there is only one matter component);
\begin{equation}
     T^{a}{}_{b} = \Psi \delta^a{}_b + 
                       n^a \mu_b \ .
\label{stressen}
\end{equation}
In this case we have $n^a = n u^a$ and $\mu_a = \mu u_a$ so the expression for the generalised pressure leads to
\begin{equation}
    \Psi = \Lambda + n\mu \ .
\end{equation}
We then have
\begin{equation}
    T^a_{\ b} = \left( \Lambda + n\mu \right) \delta^a_b + n\mu \, u^a u_b \ ,
\end{equation}
and an observer moving along with $u^a$ would  measure the energy 
\begin{equation}
    \varepsilon = u_a u^b T^a_{\ b} = -\Lambda \ .
\end{equation}
This then leads to
\begin{equation}
    n\mu = \Psi+\varepsilon \ , 
    \label{gib}
\end{equation}
and a comparison with the standard Gibbs relation from thermodynamics supports the interpretation of $\Psi$ as the pressure, $p$, in this case. This then leads to an interpretation of \eqref{presdef} as a generalised Gibbs relation, appropriate for the multifluid setting and analogous to (more phenomenological) relations introduced in the context of extended irreversible thermodynamics \cite{joubook}.

\section{Linear drift argument}

At this point it should be clear that the multifluid model extends the familiar fluid logic and introduces  features that we need to get used to. On the one hand, this is important as it allows us to describe the additional dynamical degrees of freedom of a given problem---and the coupling between them. On the other hand, the model may be too general for many situations of interest. In particular, we have to keep track of the individual  fluxes, or---if we introduce a preferred observer---the different relative velocities and the associated Lorentz factors. We may also not have a good handle on the  parameters we need to complete the model. In particular, the nuclear physics calculations that provide state-of-the-art neutron star equations of state tend to assume that the matter is in equilibrium. There are two aspects to this. First, we may insist on chemical and thermodynamical equilibrium, as appropriate if we want to determine a static matter configuration. If we want to model a dynamical scenario---even for a single matter component---we need to allow for the matter not being in equilibrium, e.g. due to changes in the local temperature. This is regularly done in modern numerical relativity simulations. The multifluid problem adds a different dimension, as the relative motion between the different components means that there is no ``obvious'' matter frame with respect to which we can develop the thermodynamics\footnote{The issue of equilibrium is intricate. For a real system, we may consider different equilibrium states, e.g. dynamical, chemical and thermodynamical. Strictly speaking, one might expect the true equilibrium to account for all these aspects (see the recent discussion in \cite{2020arXiv200800945C}). However, there are exceptions, like perfect superfluids where one may define an equilibrium state with a relative flow (see for example \cite{2020CQGra..37b5014G}). To some extent, the linear drift argument allows us to avoid this issue. }.  

In order to explain this point, and as a step towards a simplified model, let us introduce an observer, associated with a four velocity $U^a$
such that (with the speed of light $c=1$)
\begin{equation}
    u_\x^a = \gamma_\x (U^a + V_\x^a) \ ,
\end{equation}
with
\begin{equation}
    U_a V_\x^a = 0 \qquad \mbox{and where}\qquad \gamma_\x = \left(1-V_\x^2 \right)^{-1/2} \ ,
\end{equation}
provides the Lorentz factor associated with the different flows in the system.
We then have
\begin{equation}
    n_\x^a = n_\x \gamma_\x  (U^a + V_\x^a) \ ,
\end{equation}
and it is natural to introduce the number density measured by $U^a$ as 
\begin{equation}
    \hat n_\x = -U_a n_\x^a = n_\x \gamma_\x \ .
\end{equation} 
For the momenta we  get
\begin{equation}
    \mu^\x_a =  g_{ab} \left[  \hat \mu_\x  (U^b + V_\x^b)  + \sum_{\y \neq \x} \A^{\X \Y}   \hat n_\Y  W_{\y\x}^b
                \right] \ , 
\end{equation}
with the relative velocity
\begin{equation}
W_{\y\x}^a =  V_\y^a - V_\x^a \ ,
\end{equation}
and $\hat \mu_\x=-U^a\mu^\x_a$   the chemical potential measured by the chosen observer. 

The energy measured by the observer is (recall \eqref{stressen})
\begin{equation}
\hat \varepsilon = U_a U^b T^a_{\ b} = - \Psi + \sum_\x \left( U_a n_\x^a \right) \left( U^b \mu_b^\x\right) 
= - \Psi + \sum_\x \hat n_\x \hat \mu_\x \ .
\end{equation}
This seems like the intuitive generalisation of \eqref{gib}, but if we combine the result with \eqref{presdef} we see that we can no longer identify $-\Lambda$ with the measured energy. Instead, we need 
\begin{equation}
    \Psi = \Lambda + \sum_\x \hat n_\x \hat \mu_\x -   \sum_\x \hat n_\x V_\x^a \left(\hat \mu_\x V^\x_a  + \sum_{\y\neq \x} \A^{\x\y}  \hat n_\y W^{\y\x}_a \right)\ ,
\end{equation}
which leads to
\begin{multline}
    \hat \varepsilon = -\Lambda + \sum_\x \hat n_\x \hat \mu_\x V_\x^2 +  \sum_\x \sum_{\y\neq\x}   \A^{\x\y} \hat n_\x \hat  n_\y  V_\x^a \left( V^\y_a -  V^\x_a  \right) \\
    =  -\Lambda + \sum_\x \hat n_\x \hat \mu_\x V_\x^2 - {1\over 2} \sum_\x \sum_{\y\neq\x}   \A^{\x\y} \hat n_\x \hat  n_\y W_{\y\x}^2\ .
\end{multline}
The quadratic terms complicate the thermodynamics (essentially representing kinetic energy contributions relative to the chosen frame). In order to make progress we need to keep track of the entrainment and the relative velocities. From a formal point of view, we have everything we need to solve the problem. In particular, the prescription is Lorentz invariant and as such it is expected to prevent causality violations. However, if we want to make contact with a given model for the microphysics then we have a problem. The relative flows (obviously) impact on the energy of the system and this, in turn, affects the thermodynamics and issues like chemical equilibrium. If the chemical potentials are observer dependent---as they have to be in the general case---then we need to decide how to implement the notion of equilibrium \cite{2020arXiv200800945C}. Who measures what? This problem does not arise in  standard approaches to dissipative fluid dynamics, which tend to be based on formal expansions in the deviation from equilibrium \cite{roma1,roma2,baier} (see \cite{2019PhRvD.100j4020B,2019JHEP...10..034K,2020arXiv200609843G} for recent progress). Our problem is different: We need to reconcile the nonlinear multifluid model with our ``incomplete'' description of the thermodynamics \cite{carter91,2015CQGra..32g5008A,2017CQGra..34l5001A}.

As a  step towards applications, we may simplify the model by linearising in the relative flows---essentially neglecting quadratic terms in $V_\x^a$ \cite{2017CQGra..34l5002A}. This may seem like a drastic move, but it is a natural assumption because many of the situations we are interested in involve a gentle and gradual drift of one component relative to another rather than a vigorous relativistic stream. Of course, we have to be careful, because if we linearise the model we ignore the relative Lorentz factors, and we have 
\begin{equation}
    u_\x^a \approx U^a + V_\x^a \ .
\end{equation}
In effect, the model is no longer Lorentz invariant and we must not extend it into the nonlinear regime. However, if we take appropriate care in this respect, the linear drift model has a number of attractive features. In particular, it is easy to see that the connection with thermodynamics becomes straightforward. As $\gamma_\x\approx 1$, all linearly related observers  agree on the number densities and chemical potentials. We also get $\hat \varepsilon \approx \varepsilon \approx - \Lambda$ (in essence, we can drop the hats on all the scalar quantities) and the usual Gibbs relation applies. This is quite attractive.

Given the linear drift assumption, the equations of motion  \eqref{forcex} simplify to (after a bit of algebra and noting that the force only has three independent components---recall that $n^a_\x f^\x_a=0$ follows from \eqref{forcex})---so we focus on the projection orthogonal to $U^a$)
\begin{multline}
    f^\x_b = 0 \ \Longrightarrow \ \mu_\x \left( U^a \nabla_a U_b + V_\x^a \nabla_a U_b +\perp^a_b U^c \nabla_c V^\x_a \right) + \perp^a_b \nabla_a \mu_\x + V_b^\x U^a \nabla_a \mu_\x \\
    + 2U^a \sum_{\y\neq\x} \nabla_{[a} n_\y \A^{\x\y} W^{\y\x}_{b]} = 0 \ ,
    \label{fxind}
\end{multline}
where
\begin{equation}
    \perp^a_b = \delta^a_b + U^a U_b \ .
\end{equation}

For the stress-energy tensor we get (noting that the entrainment contributions cancel!)
\begin{equation}
  T^{ab} = \varepsilon U^a U^b + p \perp^{ab} + 2 \sum_\x n_\x \mu_\x U^{(a} V_\x^{b)} \ ,
\end{equation}
where we have used the usual multi-component  Gibbs relation to define the pressure
\begin{equation}
    p + \varepsilon = \sum_\x n_\x \mu_\x \ .
\end{equation}
The result now looks much more familiar. If we define the relative momentum flux 
\begin{equation}
    q^a = \sum_\x n_\x \mu_\x V_\x^a \ ,
\end{equation}
we get the final expression
\begin{equation}
  T^{a}{}_{b} \approx  \varepsilon U^a U_b + \perp^a_b p+ 2 U^{(a} q^{b)} \ .
  \label{stressheat}
\end{equation}
We recognise this as the stress-energy tensor for a system with a linearised heat flux \cite{2007LRR....10....1A}. Notably, the ``stress'' terms associated with relative flows are removed by the linearisation.

In order to complete the description, we need the continuity equations. It is easy to see that these become
\begin{equation}
    \left( U^a + V_\x^a \right)\nabla_a n_\x + n_\x \nabla_a  \left( U^a + V_\x^a \right) = 0 \ .
    \label{lincont}
\end{equation}
With this we have all the equations we need to consider specific applications, notably without at this point having committed ourselves to a specific observer. We could carry on working with a general model, but this might get confusing so it is better to consider specific examples. We will do this in three steps. First we consider a superfluid system with two components, then we turn to a problem with heat flux and a non-conserved entropy current and finally we outline a model for charged systems with resistivity. These problems have all been discussed in the literature (see \cite{2017CQGra..34l5002A} for a closely related discussion), so we do not expect to learn anything ``new'' here. Rather, the point of the discussion is to highlight the common features of the different models and consider the framework required for more complex situations step by step.

\section{Superfluids}

The variational multifluid model provides a natural framework for discussing superfluid 
systems, both in the laboratory context  and for neutron stars.  The minimum requirement of a workable model is that it should represent the expected two-fluid dynamics (e.g. the second sound and the presence of oscillation modes that reflect the two degrees of freedom, see \cite{universe7010017} for a recent discussion) and the anticipated entrainment effect (which encodes the relative ``mobility'' of the two components). As we will see, the variational model easily satisfies these criteria\footnote{In fact, the first relativistic study of neutron star oscillations \cite{comer99:_quasimodes_sf} was carried out within this framework.}. Moreover, we can go further and consider  the quantisation 
of vorticity---required for a superfluid to exhibit bulk rotation---which is easily imposed on the canonical momentum \cite{carter95:_kalb_ramond} and the implications for the 
dynamics become quite intuitive. Alternative descriptions are, of course, available (see for example the seminal contributions by Israel \cite{1981PhLA...86...79I,1982PhLA...92...77I,1985qugr.conf..246I} or the more recent discussion in \cite{carter92:_momen_vortic_helic} for the relativistic case or \cite{universe7010017} for a recent summary of the corresponding Newtonian problem). We will only comment on a few  of the relevant aspects here.

First of all, note that \eqref{forcex} is automatically satisfied if the momentum is given by the gradient of a scalar phase, $\mu^\x_a = \nabla \phi_\x$.  This implies that the fluid is irrotational\footnote{Note that the superfluid vorticity is associated with the momentum rather than the velocity of the fluid. Hence, the notion of an irrotational superfluid is subtly different from the usual notion in fluid dynamics.}. However, for most problems of interest one assumes an averaged vorticity such that the normal fluid equations apply. The argument comes with the caveat that the required (quantised) vortices may introduce new features (like mutual friction and vortex elasticity \cite{2016CQGra..33x5010A,Andersson_2020}), but here we will focus on the simplest model for bulk dynamics. The entrainment then plays a central role, so it is natural to start by providing an interpretation of this effect. 

In order to make the example clear, let us focus on the two-fluid model for neutron star cores, with a neutron superfluid ($\x=\n$) coupled to a charge-neutral conglomerate of protons and electrons ($\x=\p$). We then have
the spatial components of the proton momentum (relative to $U^a$)
\be
\mu_\p^i = \mu_\p V_\p^i +  n_\n \A^{\n\p} W_{\n\p}^i \ .
\ee
Suppose we let the observer ride along with the neutrons, in such a way that $V_\n^i=0$, then we can introduce the effective proton mass as
\be
\mu_\p^i = \left( \mu_\p - n_\n \A^{\n\p} \right) V_\p^i \equiv m_\p^* V_\p^i \ ,
\label{proteff}
\ee
 (this argument is analogous to the Newtonian discussion in \cite{universe7010017}, see also \cite{prix04:_multi_fluid}). A similar argument for the neutrons defines $m_\n^\ast$ and we see that we have
\be
\A^{\n\p} = {1\over n_\p} \left( \mu_\n - m_\n^\ast\right) =  {1\over n_\n} \left( \mu_\p - m_\p^\ast\right)  \ .
\label{neuteff}
\ee
In a neutron star core, the entrainment accounts for an important aspect of the strong interaction; the flow of neutrons pulls some protons along (and vice versa) \cite{comer03:_rel_ent}. The notion of entrainment is also important for the dynamics of the star's inner crust, in this case due to Bragg scattering off of the elastic lattice of nuclei \cite{crent1,crent6}. However, the presence of the nuclear lattice brings in additional aspects which we will not go into here, see \cite{universe7010017} for a brief summary.
Also, 
in the general case, with a number of distinct flows (or several entrainment mechanisms), the expression for the effective mass is not as simple as \eqref{proteff}---we end up with an effective mass matrix---but the concept still makes sense. 

Turning to the dynamics of the system, an advantage of considering superfluids is that both nuclear reactions and resistive scattering are suppressed, so we can focus on the non-dissipative problem with conserved fluxes. This is precisely the setting for which the variational approach was intended \cite{2007LRR....10....1A}, so we can make direct use of the equations we have written down. Moreover, in many situations of interest the relative flow between the components is small so the linear drift assumption should apply. 

When we consider the equations of motion, we need to make further decisions. We can introduce a specific observer and  we may opt to work with different combinations of the equations. It is worth explaining both of these points as different strategies have been used in the literature and it is helpful to understand how they are related.

First consider the momentum equations \eqref{fxind}. We can decide to represent the two dynamical degrees of freedom by the equations for $f^\n_a$ and $f^\p_a$. This makes the mathematics fairly ``symmetric'' in that the two fluids are treated on the same footing. As an example, this strategy has been used in the context of neutron star seismology \cite{comer99:_quasimodes_sf,2002PhRvD..66j4002A,2008PhRvD..78h3008L}. However, this approach does not exactly reflect the anticipated dynamics. It is easy to argue that the natural degrees of freedom are better described in terms of the total flux and the difference between the velocities (see \cite{universe7010017} for a discussion and pointers to the literature). The decoupling is not complete (apart from in idealised situations) but we can opt to work with a set of equations that is closer to the physics. This could, for example, involve the sum of the two force equations---which we already know is equivalent to the vanishing   divergence of the stress-energy tensor---and a weighted difference (such that the four acceleration term in \eqref{fxind} is removed). An example of this procedure, in the context of charged plasmas, can be found in \cite{2012PhRvD..86d3002A}. A somewhat simpler option is to combine the equations from the  stress-energy tensor with one of the fluid force equations to remove the four acceleration from the latter to get a direct expression for the evolution of the chosen velocity $V_\x^a$. We will outline this approach, which is close in spirit to the more ``traditional'' approach to superfluid dynamics---notably championed by Gusakov and collaborators in the context of relativity \cite{2006MNRAS.372.1776G,2011PhRvD..83j3008K}---in the following.  

Before we work out the relevant details, it is useful to consider the choice of observer. First we note that, if we add the continuity equations \eqref{lincont} and define
\begin{equation}
    n V^a = \sum_\x n_\x V_\x^a\ ,
\end{equation}
with 
\begin{equation}
    n = \sum_\x n_\x \ ,
\end{equation}
then we get
\begin{equation}
\left( U^a + V^a \right) \nabla_a n + n \nabla_a \left( U^a + V^a \right) = 0  \ .
\label{sumcont}
\end{equation}
We can clearly simplify this by letting the observer be such that $V^a=0$. This has the advantage that \eqref{sumcont} reduces to the usual single-fluid continuity equation, involving only $n$ and $U^a$. In the two-component example we are considering at the moment, $n=n_\n+n_\p$ represents the total baryon number density and we have
\begin{equation}
V^a=0 \ \Longrightarrow \ V_\p^a = - {n_\n \over n_\p} V_\n^a \ \Longrightarrow\ W_{\n\p}^a = {n\over n_\p} V_\n^a \ .
\label{eckart}
\end{equation}
This relation allows us to work with equations for $U^a$ and $V_\n^a$. The specific observer choice is analogous to the Eckart frame familiar from discussions of dissipative relativistic fluid dynamics \cite{eckart40:_rel_diss_fluid}.  

Another option is to focus on the stress-energy tensor and simplify life by making it as close to the perfect fluid as we can. This involves introducing an observer such that 
\begin{equation}
    q^a = 0 \ \Longrightarrow \ V_\p^a = - {n_\n \mu_\n \over n_\p \mu_\p} V_\n^a \ \Longrightarrow\ W_{\n\p}^a =  {p + \varepsilon\over n_\p \mu_\p} V_\n^a  \ .
    \label{landau}
\end{equation}
This choice leads to the well-known Landau-Lifshitz frame \cite{landau59:_fluid_mech}.
Other combinations are, of course, available. The bottom line is that we can use the observer to ``simplify'' some aspects of the models (possibly at the cost of complicating others...). It is also worth keeping in mind that the remaining relative velocity, $V_\n^a$, refers to different observers  and therefore has slightly different ``meaning'' in the two cases.

Suppose we commit to one of these choices, say, \eqref{landau}. Then we have
the text-book equations of motion for a perfect fluid (as the stress-energy tensor takes the same form): 
\begin{equation}
    U^a \nabla_a \varepsilon + (p+\varepsilon) \nabla_a U^a = 0 \ ,
    \label{energy}
\end{equation}
and
\begin{equation}
    U^a \nabla_a U_b = - {1\over p+\varepsilon} \perp^a_b \nabla_a p \ .
    \label{fouracc}
\end{equation}
This is nice, but... we are  still dealing with a two-fluid problem. Representing the second degree of freedom by the superfluid neutrons, we also need   
\begin{multline}
    U^a \nabla_a U_b + V_\n^a \nabla_a U_b +\perp^a_b U^c \nabla_c V^\n_a   \\+  {1\over \mu_\n} \left( \perp^a_b \nabla_a \mu_\n + V_b^\n U^a \nabla_a \mu_\n 
    - 2  U^a  \nabla_{[a} {p+\varepsilon\over \mu_\p}  \A^{\n\p} V^{\n}_{b]} \right) = 0 
  \ .  \label{fxn}
\end{multline}
Using \eqref{fouracc} to remove the four acceleration from this equation, we have (admittedly after some algebra...) an evolution equation for the superfluid component 
\begin{multline}
    \perp^a_b U^c \nabla_c \left[\left(  \mu_\n - {p+\varepsilon\over \mu_\p}  \A^{\n\p} \right) V^\n_a \right] +  \mu_\n V_\n^a \nabla_a U_b   +   \perp^a_b \left( \nabla_a \mu_\n - {\mu_\n\over p+\varepsilon}  \nabla_a p \right)  \\
    -   {p+\varepsilon\over \mu_\p}  \A^{\n\p} V_{\n}^{a}  \perp^c_b \nabla_{c} U_a = 0 \ .
\end{multline}
Recalling the effective mass from \eqref{neuteff}, we have
\begin{multline}
  \mu_\n - {p+\varepsilon\over \mu_\p}  \A^{\n\p} = \mu_\n - {p+\varepsilon\over n_\p \mu_\p}  \left(  \mu_\n - m_\n^\ast\right)   \\
 =  \mu_\n - \left( 1 + {n_\n \mu_\n \over n_\p \mu_\p} \right)   \left(  \mu_\n - m_\n^\ast\right) = m_\n^* - {n_\n \mu_\n \over n_\p \mu_\p}    \left(  \mu_\n - m_\n^\ast\right)\ , 
\end{multline}
which illustrates how entrainment impacts on the inertia of the superfluid.

It is now clear that the observer choice simplifying \eqref{fouracc} is a mirage. We still have to deal with the second degree of freedom and the relevant momentum equation has the additional features we expect, like entrainment. 
The fact that the problem remains complex is also apparent when we consider the continuity equations. We can take one of the two relations to be represented by \eqref{energy} (representing energy conservation), which is nice, but we need to keep in mind that the model requires us to keep track of two number densities, some combination of $n$, $n_\n$ and $n_\p$ or some function of them, like $\varepsilon$. In the single fluid case, we know that 
\begin{equation}
    \mu = {d\varepsilon \over dn} \ ,
\end{equation}
so \eqref{energy} can be written
\begin{equation}
\mu \nabla_a (nU^a) = 0  \ ,
\end{equation}
and the conserved flux implies the energy equation (and the other way around). This does not follow in the two-fluid case where---at the linear drift level---we have
\begin{equation}
    \nabla_a \varepsilon = \mu_\n \nabla_a n_\n + \mu_\p \nabla_a n_\p \ ,
\end{equation}
and it follows that, if we choose to work with $n$ and $n_\n$ (say) we get
\begin{equation}
    \mu_\p \nabla_a (nU^a) + ( \mu_\n - \mu_\p) \nabla_a ( n_\n U^a ) = 0   \ .
\end{equation}
This form is instructive because the second term vanishes if we impose chemical equilibrium\footnote{Remember that the ``proton'' component here includes the electrons, so the condition for beta equilibrium becomes $\mu_\n=\mu_\p$.}, bringing us back to the single fluid result. Of course, in the case of superfluids the nuclear reactions required to establish equilibrium are suppressed so this argument does not really help us.  We need to keep track of the second number density, which (in this example) is governed by 
\begin{equation}
    \left( U^a + V_\n^a\right) \nabla_a n_\n + n_\n \nabla_a  \left( U^a + V_\n^a\right) = 0 \ .
\end{equation}

Although there are many issues one might want to add to the discussion, especially regarding vortices \cite{carter95:_kalb_ramond,2016PhRvD..93f4033G,Andersson_2020,2020arXiv201210288G}, will move on at this point. The key aim was to illustrate how the multifluid model can be adapted to account for the superfluid degree of freedom and how this introduces new microphysics parameters associated with entrainment (and which have to be provided from nuclear physics \cite{comer03:_rel_ent}, see \cite{crent4, crent7} for recent efforts in this direction). The simple fact that the two-fluid equations that result involve elements of choice is important to keep in mind. 

\section{The flow of heat}

Staying at the formal level, let us turn to a problem that allows us to discuss how the model changes when the fluxes are not conserved. The simplest reasonable problem in this direction is that of heat flowing relative to matter \cite{carter83:_in_random_walk,carter_heat,andersson10:_caus_heat,2011RSPSA.467..738L,AL11}. In this case, the minimal model includes a distinct heat flow and the connection to the second law of thermodynamics (entropy must not decrease) which makes the problem dissipative. This situation can (again) be represented by two components. We will take them to be the matter ($\x=\n$) and the entropy ($\s$) and it would usually be the case that the associated heat flux is a gentle drift so we should be able to make progress within the linear approximation. 

A key aspect of the heat-flux problem is that the entropy current is not conserved. Letting $s^a = n_\s^a$ we have (for an isolated system)
\begin{equation}
    \nabla_a s^a = \Gamma_\s \ge 0 \ ,
    \label{entflux}
\end{equation}
in accordance with the second law of thermodynamics\footnote{We will make the common assumption that the second law is to be imposed at the local level, even though it is (strictly) a global statement relating to the total entropy of the system.}. Given this, let us first ask a general question. What happens to the multifluid formalism if the fluxes are not conserved? This forces us to either adjust the variational argument (as explored in \cite{2020arXiv200800945C,2015CQGra..32g5008A,2017CQGra..34l5001A}) or proceed in a more phenomenological manner. Here, we will take the latter approach, essentially adding the main dissipation mechanism of interest and not worrying too much about the general problem. We simply note that if we have
\begin{equation}
    \nabla_a n_\x^a = \Gamma_\x \ ,
\end{equation}
in \eqref{divT}, then we must adjust the force equation in such a way that \cite{2015CQGra..32g5008A,carter91}
\be
2 n_\x^b  \nabla_{[b}  \mu^\x_{a]} + \Gamma_\x \mu^\X_a =  R^\X_a \ ,
\label{momeq}
\ee
with
\be
\sum_\x R^\x_a = 0 \ .
\label{Rcon}
\ee
and the constraint
\be
\Gamma_\x = - {1\over  \mu_\x} \left( u_\x^a R^\x_a \right) \ . 
\label{gamdef}
\ee
That is, the reaction rate determines the time component (in a co-moving frame) of the resistivity  $R^\x_a$ (which, in turn, must be included in \eqref{momeq} since the original force term is orthogonal to $n_\x^a$).

Let us now focus on a problem involving a single matter component---letting $n_\n^a = n^a$, for simplicity, with the corresponding chemical potential being $\mu_\n =\mu$---and an entropy component (from before) for which the chemical potential is the temperature, so $\mu_\s=T$. At the linear drift level, we then have the (familiar) thermodynamical relation
\begin{equation}
p+\varepsilon = n\mu + sT \ .
\end{equation}
It also follows that \eqref{entflux} becomes
\begin{equation}
    U^a \nabla_a s + s \nabla_a U + \nabla_a \left( {q^a \over T} \right) = \Gamma_\s \ ,
\end{equation}
where we have introduced the heat flux (in the usual way)
\begin{equation}
    q^a = sT V^a_\s \ .
\end{equation}
Turning to the momentum of the thermal component, when we allow for entrainment between matter and heat---this may seem like an unusual idea, but let us go along with it and see where it takes us---we have
\begin{equation}
\mu^\s_a = T \left( U_a + V^\s_a \right) + \A^{\s\n}n W^{\n\s}_a \ ,
\end{equation}
and we may introduce the effective mass for the thermal component through
\begin{equation}
    \A^{\n\s} = {1\over n} \left( T-m_\s^*\right) \ .
\end{equation}
This relates the entropy entrainment to the effective inertia of heat, a notion that is expected to play an important role in the relativistic heat problem \cite{1997CQGra..14.2239H,1998CQGra..15..407H}.

So far, the results are straight translations of previous relations. Moreover, 
opting---as in the superfluid case---to work in the centre of momentum frame, we have
\begin{equation}
    W_{\s\n}^a = {p+\varepsilon \over n\mu} V_\s^a \approx {q^a\over sT}
\ , \end{equation}
(assuming that $sT\ll n\mu$, which seems likely to be the case in most situations of interest).
This means that we have
\begin{equation}
\mu^\s_a \approx TU_a + \left[T - { \left( p+\varepsilon \right) \A^{\s\n} \over \mu}  \right] {q_a \over sT} \approx T U_a + {m_\s^* \over sT} q_a \ .
\end{equation}

In order to proceed, we need an expression for the resistivity, $R_\x^a$. Making use of the results from \cite{2017CQGra..34l5001A} we have---assuming that the reaction rates $\Gamma_\x$ and the resistivity coefficients $\mathcal R^{\x\y}$ are provided by the microphysics, 
\be
R^\x_a =  \Gamma_\x  \mu_\x u^\x_a +  \sum_{\y\neq\x} \mathcal R^{\x\y} (\delta_a^b + V_\x^bU_a)  W^{\y\x}_b  \ ,
\label{pheno}
\ee
for all material particles, along with the constraint on the entropy:
\be
R^\s_a = - \sum_{\x\neq\s} R^\x_a \ .
\label{Rscon}
\ee
In the present context, with only two components and $\Gamma_\n=0$, this leads to 
\begin{equation}
T \Gamma_\s = - u_\s^a R^\s_a \approx (U^a + V_\s^a)  \sum_{\x\neq\s} R^\x_a = \mathcal R^{\n\s} W_{\n\s}^2 \ge 0 \ .
\label{TGs}
\end{equation}
We see that the $\mathcal R^{\n\s} = \mathcal R^{\s\n}$ coefficient is required to be positive by the second law of thermodynamics. 

It is also useful to make contact with the standard discussion of the problem by introducing the thermal conductivity $\kappa$ such that
\begin{equation}
    \Gamma_\s = {q^2 \over \kappa T^2} \ \Longrightarrow \ \mathcal R^{\s\n} \approx {s^2 T\over \kappa} \ .
\end{equation}
It follows that---as should have been expected---the entropy creation rate $\Gamma_\s$ is quadratic in the relative velocity. Hence, the corresponding contribution to the momentum equations should be neglected at the linear drift level. This leaves only the resistive contribution to the entropy momentum equation. In effect, we only need to add a term to the right-hand side of \eqref{fxind} (where we need to keep in mind that we divided through by an overall factor of $n_\x$) 
\begin{equation}
{1\over s} \perp^a_b \ R^\s_a = {1\over s} \perp^a_b  \left[ \Gamma_\s  T (U_a + V^\s_a) +   \mathcal R^{\s\n} (\delta_a^c + V_\s^b U_a)  W^{\n\s}_c  \right] \approx {1 \over s} \mathcal R^{\s\n} W^{\n\s}_b = - {q_b \over \kappa} \ .
\end{equation}
This means that we have an evolution equation for the heat flux:
\begin{equation}
    \perp^a_b U^c \nabla_c \left( {m_\s^* \over sT} q_a \right) + {q^a \over s} \nabla_a U_b  + {m_\s^* - T \over sT} q^a \perp^c_b \nabla_c U_a +  {1\over \kappa} q_b  = - \perp^a_b \left( \nabla_a T - {T\over p+\varepsilon} \nabla_ a p \right) \ .
\end{equation}

This model---which arises naturally in the multifluid framework---helps explain the main features of the relativistic heat problem. In the absence of heat flow the terms on the right-hand side bring out the classic result (when combined with the Tolman-Oppenheimer-Volkoff equations for hydrostatic equilibrium) that the redshifted temperature is uniform \cite{1930PhRv...35..904T}. In effect, heat has weight. If we neglect the terms involving $U^a$, we get the relation for the heat flux $q^a$ required for a derivation of the equations used to describe the gradual cooling of neutron stars \cite{2007LRR....10....1A}. In the general case, we have a telegraph-type equation---as in extended irreversible thermodynamics \cite{joubook}---with the effective mass $m_\s^*$  regulating the thermal relaxation, which is key to keeping the model causal and stable \cite{AL11}. Perhaps the main lesson of the exercise is that it paid off to take the entropy entrainment seriously.

\section{Electromagnetism}

The third problem we are going to consider, again from the two-fluid point of view, introduces charged flows and the coupling to electromagnetism. The marriage between electromagnetism and multifluid model is easy and comfortable \cite{2012PhRvD..86d3002A}, as should be expected given that both derive from an action principle \cite{2007LRR....10....1A}. When we consider the problem of charged flows in relativity, with the aim to reach beyond  text-book results, a minimal requirement is to add resistivity to magnetohydrodynamics \cite{bekor,pale,2013PhRvD..88d4020D,2020MNRAS.491.5510W}. Essentially, we want to write down a version of Ohm's law that connects to the notion of the charge current as arising from the relative flow between two components \cite{kandus,koide09}.  

Starting from the usual assumption of minimal coupling, the electromagnetic Lagrangian is built from the anti-symmetric Faraday tensor;
\be
F_{ab} = 2 \nabla_{[a} A_{b]}  \ ,
\ee 
where $A_a$ is the vector potential, 
and the  electromagnetic field couples to the matter flow through the charge current $j^a$ (via a coupling term $\mu_0 j^a A_a$, where the coupling constant $\mu_0$ is the magnetic permeability). In order for this construction to be gauge invariant, the current must be 
conserved. That is, if the model involves a number of charge carriers, each with a charge $q_\x$ per particle, we have the constraint
\be
\nabla_a j^a = 0  \ ,
\ee
where
\begin{equation}
    j^a = \sum_\x j_\x^a = \sum_\x q_\x n_\x^a \ .
\end{equation}
Clearly, this constraint is automatically satisfied when the individual fluxes are conserved. If we account for reactions, as we may want to do for neutron stars,  we must impose overall charge conservation
\be
\sum_\x q_\x \Gamma_\x = 0 \ .
\label{gaugecon}
\ee

The coupling to the vector potential impacts on the matter momentum in the fashion anticipated from text-book quantum mechanism, and we have
\be
\bar \mu^\x_a = \mu^\x_a + q_\x A_a \ .
\ee
Formally, this is the only change we need to make to the fluid equations.
As long as  we change $\mu^\x_a \to \bar \mu^\x_a$ the individual force equations \eqref{forcex} remain unchanged. Meanwhile, a variation of the action with respect to the vector potential (keeping $j^a$ fixed!), leads to the Maxwell equations
\be
\nabla_b F^{ab} = \mu_0 j^a \ ,
\label{max1}
\ee
 completed by the symmetry relation
\be
\nabla_{[a}F_{bc]}=0 \ .
\label{max2}
\ee
Finally, a variation with respect to the spacetime metric leads to the electromagnetic contribution to the stress-energy tensor, which satisfies
\be
\nabla_a T^{ab}_\mathrm{EM} =  j_a F^{ab} \equiv - f_\mathrm{L}^b \ ,
\label{loren}\ee
 in turn, defining the Lorentz force, $f_\mathrm{L}^b$. 
 Basically, we may take the view that the matter stress-energy tensor still takes the form from \eqref{tab} (in terms of the pure matter momenta) and that we have\footnote{This separation follows from the assumption of minimal coupling. It would not be possible in a general polarised medium, where the presence of the electromagnetic field affects the matter properties.}
\be
\nabla_a T^{ab}_\mathrm{M} = -  \nabla_a T^{ab}_\mathrm{EM} = - j_a F^{ab} = f_\mathrm{L}^b \ ,
\label{TMeq}
\ee
If we insist on separating the electromagnetic contributions from the matter ones in this way, we end up with a new set of momentum equations of form
\be
2 n_\x^b  \nabla_{[b} \mu^\x_{a]}  + \Gamma_\x  \mu^\x_a =  j_\x^b F_{ab}+ R^\X_a - \Gamma_\x q_\x A_a \ .
\label{momeqs}
\ee

Let us unpick the different contributions to \eqref{loren} and \eqref{momeqs}, starting with the electromagnetic terms. In order to help intuition, it may be useful to introduce the (obviously observer dependent) electric and magnetic fields, $e_a$ and $b_a$ (although it is obvious from \eqref{momeqs} that, if $\Gamma_\x\neq0$ we must also keep track of the vector potential, see \cite{baum} for the starting point for such a formulation). These are defined by
\be
F_{ab} = 2 U_{[a} e_{b]} + \epsilon_{abc} b^c  \ ,
\ee
where $\epsilon_{abc} = U^d \epsilon_{dabc}$ is associated with a right-handed tetrad moving along with the observer (and noting that each field has three independent components as $U^a e_a = U^a b_a=0$). We also have the charge current
\be
j^a = \sigma U^a + J^a  \ , \quad \mbox{with}\quad J^a U_a = 0
\ . \ee
That is, we have
\be
\sigma = - U_a j^a =  \sum_\x q_\x n_\x \ , 
\ee
and it is evident that---since they agree on the number densities---all observers within the linear drift family must agree on the charge density $\sigma$. This is important as it means that we can consistently impose the usual condition of charge neutrality\footnote{Whether this condition should be expected to hold on the different scales of the problem is a different question...} simply by setting $\sigma = 0$.
This then leads to
\be
\nabla_a J^a = 0 \ , 
\ee
while the  Lorentz force takes the form
\be 
f_\mathrm{L}^b = - j_a F^{ab} = \left(J_a e^a\right) U^b + \epsilon^{bac} J_a b_c \ .
\ee
Meanwhile, for the individual momentum equations \eqref{momeqs} we need
\begin{equation}
     j^b_\x F_{ab} = q_\x n_\x \left[ e_a + U_a \left( V_\x^b e_b \right) + \epsilon_{abc} V_\x^b b^c \right]  \ .
\end{equation}
Finally, we need the resistivity $R^\x_a$, which is given by \eqref{pheno} (with $\bar \mu^\x_a$ replacing $\mu^\x_a$). That is, in \eqref{momeqs} we need 
\begin{multline}
    R^\x_a - \Gamma_\x \bar \mu^\x_a = \sum_{\y\neq\x} \mathcal R^{\x\y}\left( \delta^b_a + V_\x^b U_a\right) W^{\y\x}_b + \Gamma_\x \left[\bar\mu_\x \left( U_a + V^\x_a\right)  - \bar \mu^\x_a \right] \\
    \approx \sum_{\y\neq\x} \left( \mathcal R^{\x\y} - n_\y \Gamma_\x  \mathcal A^{\x\y} \right) W^{\y\x}_a  - q_\x \Gamma_\x ( \perp^b_a + V^\x_a U^b) A_b \ .
\end{multline}

In order to make the discussion more specific, let us focus on the simplest neutron star setting\footnote{Focusing on the star's interior, as the exterior near-vacuum conditions are obviously different.}, i.e. a four-component system, with neutrons (n), protons (p), electrons (e) and entropy (s). This system reduces to two fluid degrees of freedom if we allow the electrons to drift relative to the other components (which are taken to be locked\footnote{This may not be a fair reflection on the actual physics since the electrons may be the main entropy carriers in the system, but we could easily change the model to either lock the entropy to the electrons or even allow for the heat to drift relative to all matter components. All that happens is that the equations get a little bit less intuitive.}), thus providing the required charge current. In effect, if we take $V_\n^a=V_\p^a=V_\s^a \neq V_\e^a$ then the charge current is given by (assuming local charge neutrality and using $q_\e = -e$, which should not be confused with the local electric field $e_a$)
\be
J^a = e n_\e W_{\p\e}^a \ \Longrightarrow \ V_\p^a = V_\e^a + {1\over e n_\e} J^a \ .
\label{Jcurr}
\ee
We also have the condition for the centre of momentum frame:
\begin{equation}
    \sum_\x n_\x \mu_\x V_\x^a = 0 \ 
    \Longrightarrow \ \left( n_\n \mu_\n + n_\p \mu_\p + sT \right) V_\p^a + n_\e \mu_\e V_\e^a = 0  \ ,
\end{equation}
where local charge neutrality leads to $n_\p = n_\e$. Putting things together, we have
\begin{equation}
    V_\e^a = - \left( 1 - {n_\e \mu_\e \over p + \varepsilon}\right) {J^a \over e n_\e } \approx - {J^a \over e n_\e } \ ,
\end{equation}
where the approximation is valid as long as $n_\e \mu_\e \ll p + \varepsilon$. We will assume that this is the case\footnote{This assumption should be (moderately) reasonable for neutron stars, but it is useful to keep in mind that $\mu_\e\gg m_\e$ in high density neutron star matter, so one has to be a little bit careful with the usual argument that the ``electron is light compared to the proton''.}.

We now have all the ingredients we need to write down the momentum equation for the electrons (for which we ignore the possible entrainment coupling\footnote{Note that we ignore the argument we just made for the relevance of entropy entrainment here, which may seem somewhat inconsistent.}, leading to $\mathcal A^{\e\x} = 0$). 
Adapting the previous results, we  get 
\begin{multline}
    \perp^a_b U^c \nabla_c (\mu _\e V^\e_a ) + \mu_\e V_\e^a \nabla_a U_b +  \perp^a_b \nabla_a \mu_\e - {\mu_\e \over p+\varepsilon} \perp^a_b \nabla_a p \\ =
    - e \left( e_b + \epsilon_{bac}V_\e^a b^c \right)
    + {1\over e n_\e^2} \sum_{\y\neq\e} \mathcal R ^{\e\y} J_b + {e \Gamma_\e \over n_\e} \left( \perp^a_b - {J_b U^a \over e n_\e} \right) A_a \ .
\end{multline}
Assuming that the electron flux is conserved ($\Gamma_\e = 0$), defining the total resistivity
\begin{equation}
    \mathcal R = \sum_{\y\neq\e} \mathcal R ^{\e\y} \ ,
\end{equation}
and introducing the electro-chemical field \cite{1983MNRAS.204.1025B,2017CQGra..34l5002A}
\be
\mathcal E^a = e^a + {1\over e} \perp^{ab} \left(  \nabla_b \mu_\e - {\mu_\e \over p+\varepsilon} \nabla_b p \right) \ ,
\label{electrochem}
\ee
we have
\begin{equation}
\mathcal E_b - {1\over e n_\e} \epsilon_{bac} J^a b^c - {\mathcal R \over e^2 n_\e^2} J_b
= \perp^a_b U^c \nabla_c \left( {\mu _\e \over e^2 n_\e}  J_a \right) + {\mu _\e \over e^2 n_\e} J^a \nabla_a U_b  \ .
\label{currdyn}
\end{equation}
In general, this dynamical equation for the charge current complements the equation for total energy-momentum conservation. If the latter is taken from \eqref{TMeq} and we work in the centre of momentum frame (as we have assumed), then $T_\mathrm{M}^{ab}$ notably has the perfect fluid form. This supports the logic often taken as starting point for relativistic magnetohydrodynamics in the literature, but it is evident that this assumption strictly depends on the choice of observer (or may only be approximately true \cite{2017CQGra..34l5002A}).

At this point, the problem still has two distinct  degrees of freedom, represented by $U^a$ and $J^a$, both governed by dynamical evolution equations. In essence, the model retains the underlying plasma properties. To simplify things, one would typically start by ignoring the right-hand side of \eqref{currdyn} (essentially the electron inertia). This leads to a version of Ohm's law:
\begin{equation}
\mathcal E_b - {1\over e n_\e} \epsilon_{bac} J^a b^c \approx  {\mathcal R \over e^2 n_\e^2} J_b \ .
\label{ohm}
\end{equation}
If we further ignore the ``battery terms'' (the gradients in \eqref{electrochem}) and the Hall term in \eqref{ohm} we are left with
\begin{equation}
    e_b \approx  \hat {\mathcal R} J_b \ ,
\end{equation}
where we have defined $\hat{\mathcal R} = \mathcal R/e^2 n_\e^2$. This result  should be familiar. 

It is obviously rewarding to see that we arrive at the expected result from the multifluid model, but the most important insights relate to the steps involved in the derivation. Depending on the degree of accuracy we require, we can undo the steps one by one and check if the assumptions are warranted for different problems of interest. 

Finally, in the ideal limit---when the resistivity vanishes so we have  a perfect conductor (with $\hat{\mathcal R} = 0$)---we see that the electric field vanishes in the fluid frame. This is the usual assumption of magnetohydrodynamics. Of course, we have to be mindful of the chosen frame and the assumption that the other contributions to \eqref{currdyn} can all be neglected. Again, the usual result is most likely only approximately true.

\section{Radiation hydrodynamics}

So far we have considered three models relating to a range of observed neutron star phenomena: pulsar glitches explored through radio timing and considered as possible gravitational-wave sources  (superfluid hydrodynamics), neutron star cooling observed by x-ray satellites (heat flux) and a range of magnetic field related phenomena, from explosive dynamics to the gradual long-term field evolution over thousands of years (resistive magnetohydrodynamics).  These examples demonstrate the versatility of the multifluid formalism, but (at least) one interesting problem remains to be contemplated: the coupling between matter and radiation. This is a crucial issue for both supernova core collapse (where neutrinos are thought to provide the key explosion mechanism) and neutron star mergers (where the merger remnant heats up to temperatures similar to those reached in high-energy colliders). The problem of neutrino transport is extremely challenging \cite{2020LRCA....6....4M}, but it has one simple and intuitive limit. At high temperatures, the neutrinos are trapped by the matter and they may have a short enough mean-free path that we can meaningfully describe them as a fluid \cite{1973erh..book.....P,1976ApJ...207..244H}. Without going into the fine print of the problem, it seems useful to ask to what extent we can expect to make progress with such a strategy. In essence, we want to consider if the radiation problem can be meaningfully considered from the multifluid perspective. 

At a glance, the idea seems promising. Suppose we consider the simple case of a single photon---the logic is similar for neutrinos, but in that case we also have to consider lepton number conservation---then we have a radiation stress-energy tensor
\begin{equation}
    T_\mathrm{R}^{ab} = \left({ h \over \nu }\right) k^a k^b \ ,
\end{equation}
where $h$ is Planck's constant, $\nu$ is the photon frequency and 
\begin{equation}
    k^a = \nu(U^a + l^a) \ ,
\end{equation}
is null, so $l_a l^a = 1$. The vector $l^a$ provides the direction of propagation relative to the observer $U^a$. It is easy to see that this leads to the anticipated energy
\begin{equation}
    \varepsilon = U_a U_b T_\mathrm{R}^{ab} = h\nu \ .
\end{equation}
We can obviously rewrite the stress-energy tensor as
\begin{equation}
    T_\mathrm{R}^{ab} = \varepsilon U^a U^b + 2U^{(a} q_\mathrm{R}^{b)} + P^{ab} \ ,
    \label{radstress}
\end{equation}
with
\begin{equation}
    q_\mathrm{R}^a = \varepsilon l^a \ ,
\end{equation}
and
\begin{equation}
    P^{ab} = \varepsilon l^a l^b \ .
\end{equation}
Viewing these results from a fluid perspective, it is clear that \eqref{radstress} reminds us of, for example, \eqref{stressheat}---after all, it has the form of a  general stress-energy tensor. The comparison is, however, misleading. For example, in the case of radiation, we do not have---at least not yet---an identifiable ``drift velocity'' that can be linearised. A photon moves at the speed of light regardless of the observer. There may be a meaningful concept of ``collective drift'', but this is not evident yet.

Formally, we can always assume minimal coupling between matter (a perfect fluid, say) and radiation in such a way that
\begin{equation}
\nabla_a T^{ab}=0 \ \Longrightarrow \ \nabla_a T_\mathrm{M}^{ab} = - \nabla_a T_\mathrm{R}^{ab} \equiv G^b \ .
\end{equation}
This is obvious, as is the fact that we run into trouble if we try to bring the variational multifluid formalism to bear on the problem.  The variational approach builds on the conservation of fluid element worldlines \cite{2007LRR....10....1A}. Because the radiation is null, this argument does not apply (in the free-streaming limit). Of course, the logic should work for neutrinos (which have a small---but nevertheless---rest mass) and one might also make progress by thinking of the radiation as associated with an effective mass (as we did in the case of heat, where the entropy is also massless, strictly speaking). Intuitively, the argument should work for systems with ``trapped'' radiation, so the objection may be more a formality than a real stumbling block. One might, for example, consider a picture where the interaction with matter (perhaps through some form of entrainment?) endows the radiation with an effective mass. This effective mass would then have to vanish in the free streaming limit (which inevitably becomes singular given that $k^a$ is null, but one might be able to make this workable). 

The real question is not if we can build a multifluid model for the radiation problem. Rather, we might want to ask if we should. In order for the effort to make sense, we would have to learn something new by going in this direction. This is where we seem to run into trouble. In a realistic setting, we are not dealing with a single photon with a well defined direction of propagation. Instead we have an energy/momentum spectrum, with a distribution typically governed by Boltzmann's equation. The formalism for this is well developed and the issues associated with it are well understood (in the context of kinetic theory). The actual challenge is computational \cite{2020LRCA....6....4M,2012PTEP.2012aA301K}, given the added dimensions associated with the radiation phase space. This is why practical implementations \cite{1992ApJS...80..819S,2013MNRAS.429.3533S,2012PThPh.127..535S,2020ApJ...900...71A} typically involve simplifying the problem by integrating out momentum aspects, leading to a well-defined moment expansion. Truncating this expansion, we arrive at a radiation stress-energy tensor of form \eqref{radstress}, which means that the formal fluid comparison ``makes sense'' \cite{2020Symm...12.1543G} but in reality we do not learn much from this. The key point is that we do not have a single relative velocity that can be used to identify the ``second fluid'' in the problem. Instead, we have a situation where different pieces (say, frequency bins) are associated with separate momentum contributions, leading to the multi-group approach to numerical simulations. We could always think of this as a souped up multifluid problem, with different ``fluids'' representing different frequency bins, but there hardly seems to be any merit to this idea. It certainly does not provide us with anything  that the current technology does not already consider. In essence then, the coupled problem matter and radiation does not appear to be a multifluid problem, at least not in a meaningful sense. There may be a strongly coupled limit where the idea applies, but this seems somewhat artificial (as we cannot model the transition to the free-streaming limit) and it does  not really help us make progress on the problem we actually need to solve. 

\section{Concluding remarks}

This survey was written with the aim of providing a (somewhat introductory) perspective on the ongoing effort to develop a flexible formalism for relativistic multifluid systems. The connection between this (more formal) endeavour and the need for a reliable description of the physics required for multimessenger astronomy is fairly clear. A number of relevant physics scenarios require multifluid aspects to be taken into account. As illustrations, we considered three  settings: i) the problem of superfluid dynamics, ii) the issue of heat flow, and iii) the combination of charge currents and resistivity leading to Ohm's law. In each case we demonstrated that the multifluid approach provides valuable insight,  motivating concepts (like the effective inertia of heat) which have often been introduced phenomenologically. The simple fact that the different problems are described within a single over-arching framework is important as it makes a combination of the problems fairly straightforward \cite{2017CQGra..34l5002A}. For example, one can easily formulate a three-fluid model to demonstrate the thermo-electric effect or consider how the onset of superfluidity impacts on the evolution of a neutron star's magnetic field. These---and many other---interesting problems are now within reach. 

This is rewarding, but the job is not done. Neutron stars are complex systems, and the multifluid model is not (quite) able to cover all the relevant aspects. Most notably, many scenarios of interest involve dissipation and we also need to improve the description of the coupling between matter and radiation (like photons or neutrinos). In this survey, we only tried to explain why the latter is not a multifluid problem (at least not in the usual sense). The argument is fairly obvious and may seem rather dismissive. This was, however, not quite the intention. The aim was to argue that the fluid approach to radiation will always be limited and to suggest that the multifluid logic is only truly useful if it brings new perspective. The three models we discussed provide such insights while the consideration of the radiation problem does not (yet) do so. There is, of course, scope for progress in this direction. We may just have to dig deeper.  

While the variational framework for multifluid systems is mature and covers a number of aspects with (more or less) immediate importance for astrophysical applications, a number of issues require further thought. This involves both formal aspects and applications. For example, on the formal side, it would be interesting to extend existing work on elasticity (see \cite{2007LRR....10....1A} for a summary) to account for plastic flow and viscoelasticity and consider the impact of the superfluid  neutron component that is present in the inner crust of a neutron star. This formal development then has an immediate application,  as one may consider the impact of the gradual superfluid decoupling (as the star cools) on the evolution of a neutron star's magnetic field. This would be naturally phrased as a three-fluid problem. Of course, we also need to consider what is calculable and what is not (given our understanding of the physics). In particular, we need to make sure that the models remain grounded in realistic microphysics. For any problem involving  neutron star crust dynamics, the entrainment \cite{crent1,crent6} is expected to play an important role but a number of related issues remain (somewhat) unsettled (see \cite{crent4,crent7}). The context of elasticity is, of course, just one of many possible future directions. The framework we have outlined should be flexible enough to cope with pretty much anything we choose to throw at it.

\section*{Funding}
The work described in this article was supported by STFC through grant number ST/R00045X/1.

\section*{Acknowledgments}
My understanding of relativistic fluid dynamics (as it is) has been developed over the last two decades or so,  making use of Isaac Newton's suggested vantage point of the elevated  ''shoulders of giants''. In particular, I have learned much of what (I think) I know from the work of Brandon Carter and Bernard Schutz.  Without their insights my perspective would be much more limited. I have also had the pleasure of working with many great collaborators. When it comes to fluids, Greg Comer stands out. I am not going to try to work out how many times he has had to explain things to me, but it is a large number... and counting. 

\section*{Data Availability Statement}
No data was used in this paper.

\bibliographystyle{frontiersinHLTH&FPHY}

\end{document}